\documentclass{paper}
\usepackage{epsfig, graphicx}
\usepackage[english]{babel}

\begin{document}

\title{Condensation of Silica Nanoparticles on a Phospholipid Membrane}
\author{\large \rm \bf V.\,E.\,Asadchikov$^a$, V.\,V.\,Volkov$^a$, Yu.\,O.\,Volkov$^a$,  }
\maketitle
\vspace{-0.2in}
\noindent {\large \rm \bf \fbox{K.\,A.\,Dembo$^a$}, I.\,V.\,Kozhevnikov$^a$, B.\,S.\,Roshchin$^a$,}  

\vspace{0.1in}
\noindent {\large \rm \bf  D.\,A.\,Frolov$^a$, and A.\,M.\,Tikhonov$^{b,}$\/\footnote{tikhonov@kapitza.ras.ru}}

\vspace{0.05in}
\leftline{$^a$\it A.\,V.\,Shubnikov Institute of Crystallography, Russian Academy of Sciences,}
\leftline{\it Leninskii pr. 59, Moscow, 119333 Russia}

\vspace{0.05in}
\leftline{$^b$\it P.\,L.\,Kapitza Institute for Physical Problems, Russian Academy of Sciences,}
\leftline{\it ul. Kosygina 2, Moscow, 119334, Russia}

\rightline{\today}

\abstract{The structure of the transient layer at the interface between air and the aqueous solution of silica nanoparticles
with the size distribution of particles that has been determined from small-angle scattering has been studied
by the X-ray reflectometry method. The reconstructed depth profile of the polarizability of the substance
indicates the presence of a structure consisting of several layers of nanoparticles with the thickness that is
more than twice as large as the thickness of the previously described structure. The adsorption of
1,2-distearoyl-sn-glycero-3-phosphocholine molecules at the hydrosol/air interface is accompanied by the condensation
of anion silica nanoparticles at the interface. This phenomenon can be qualitatively explained by the
formation of the positive surface potential due to the penetration and accumulation of Na$^+$ cations in the
phospholipid membrane.}
\vspace{0.25in}

\large
Silica sol (the solution of SiO$_2$ nanoparticles in water containing a small amount of NaOH) forms a
strongly polarized interface with air [1]. The gradient of the surface potential in this system appears due to
the difference between the potentials of the electric image forces for Na$^+$ cations and nanoparticles (macroions) 
with a large negative charge ($\sim 1000$ electrons). For this reason, the plane of the closest approach of
anion particles with the surface is $\sim 10$ nm, whereas cations are accumulated directly at the interface. It
was previously reported that unique boundary conditions make it possible to form macroscopically planar
lipid membranes on the surface of hydrosol [2]. In this work, we study the distribution of nanoparticles in a
wide transient layer at the interface with the lipid membrane of 1,2-distearoyl-sn-glycero-3-phosphocholine (DSPC)
 molecules (Fig. 1). According to our data, the adsorption of phosphocholine molecules at
the hydrosol/air boundary is accompanied by the condensation of silica nanoparticles at this interface.

The DSPC films were prepared and studied in a fluoroplastic dish with a diameter of about 100 mm
according to the method reported in [1]. Using a Hamilton 10 $\mu$l syringe, two or three drops (with the
total volume of about 10 $\mu$l) of the solution of phospholipid in chloroform ($\sim 3 \times 10^3$ mole/l) were deposited 
on the surface of the liquid substrate. The amount of the substance is enough to form no more than two
monolayers of lipid. The substrates were Ludox SM-30 (30 wt \% SiO$_2$ and 0.2 wt \% Na) and Ludox TM-50
(50 wt \% SiO$_2$ and 0.3 wt \% Na) standard solutions (Grace Davison) [3]. Drop spreading over the surface
was accompanied by a decrease in the surface tension ã of the air/hydrosol from $\sim$ 74 to $\sim$ 50 mN/m, which
was detected by the Wilhelmy method using an NIMA PS-2 surface pressure sensor. Then, the sample was
placed in equilibrium at room temperature in an air-tight cell for $\sim$ 12 h.

\begin{figure}
\hspace{0.5in}
\epsfig{file=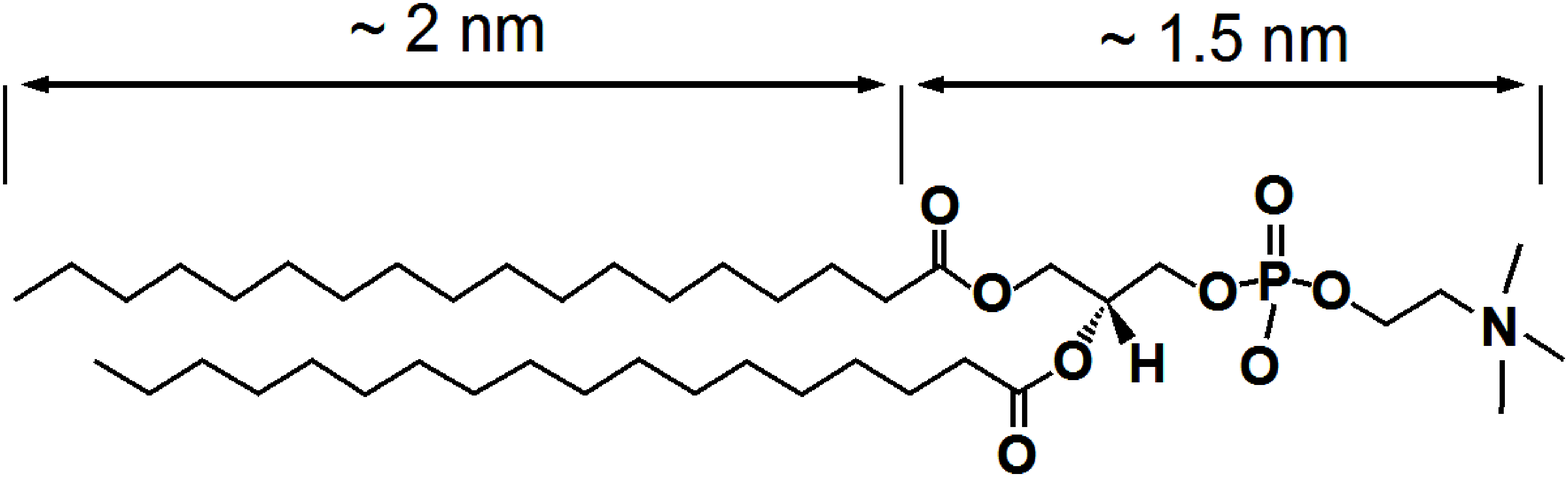, width=0.65\textwidth}

Figure 1. Molecular structure of 1,2-distearoyl-sn-glycero-3-phosphocholine.
\end{figure}

To facilitate the interpretation of the scattering data and to reveal the effect of the polydispersity in the size
distribution of particles, we preliminarily examined silica solutions by the small-angle scattering method
[4]. The bulk hydrosol sample was prepared in a glassy capillary and the measurements were performed using
an AMUR-K X-ray diffractometer [5]. Figure 2 shows the distributions obtained for particles in the SM-30
and TM-50 solutions.

The results indicate that the characteristic diameter of particles $d$ for the SM-30 solution is $\sim 10$ nm; 
this solution contains a small amount of larger particles with a size of $d \sim 20$ nm. 
The distribution of particles in the TM-50 solution exhibits a narrow peak at $d = 27$\,nm.

\begin{figure}
\hspace{0.5in}
\epsfig{file=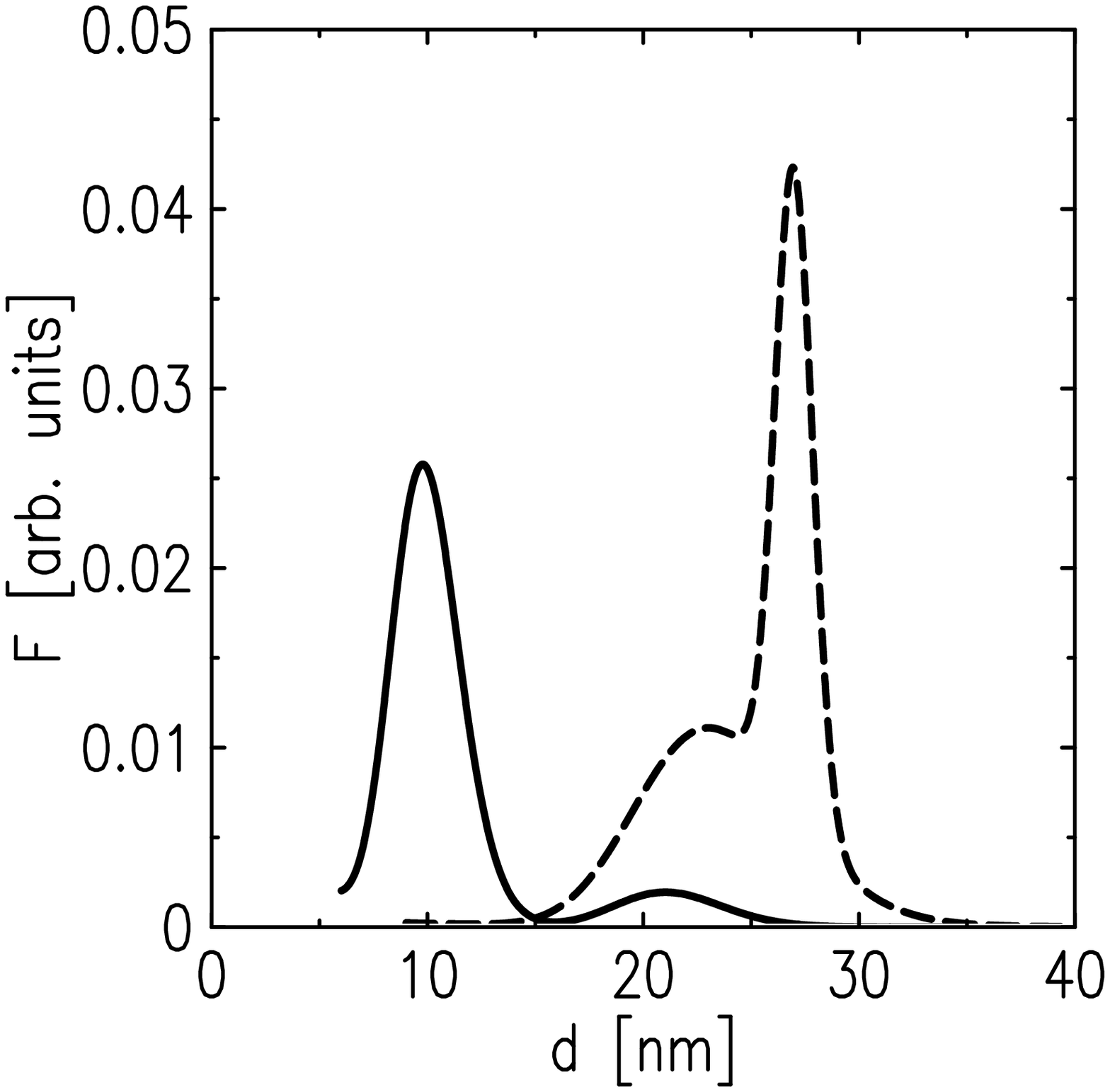, width=0.65\textwidth}

Figure 2. Diameter distribution F of particles in the (solid line) SM-30 and (dashed line) TM-50 silica solutions
according to small-angle scattering.
\end{figure}

All X-ray reflectometry experiments were performed using a multifunctional X-ray diffractometer
with a mobile tube-detector system [6]. The measurements were performed at a wavelength of $\lambda \sim 1.54$\,\AA \ \  
($\delta\lambda/\lambda \sim 10^{-4}$) at an angular resolution of $\sim 10^{-4}$ rad.

\begin{figure}
\hspace{0.5in}
\epsfig{file=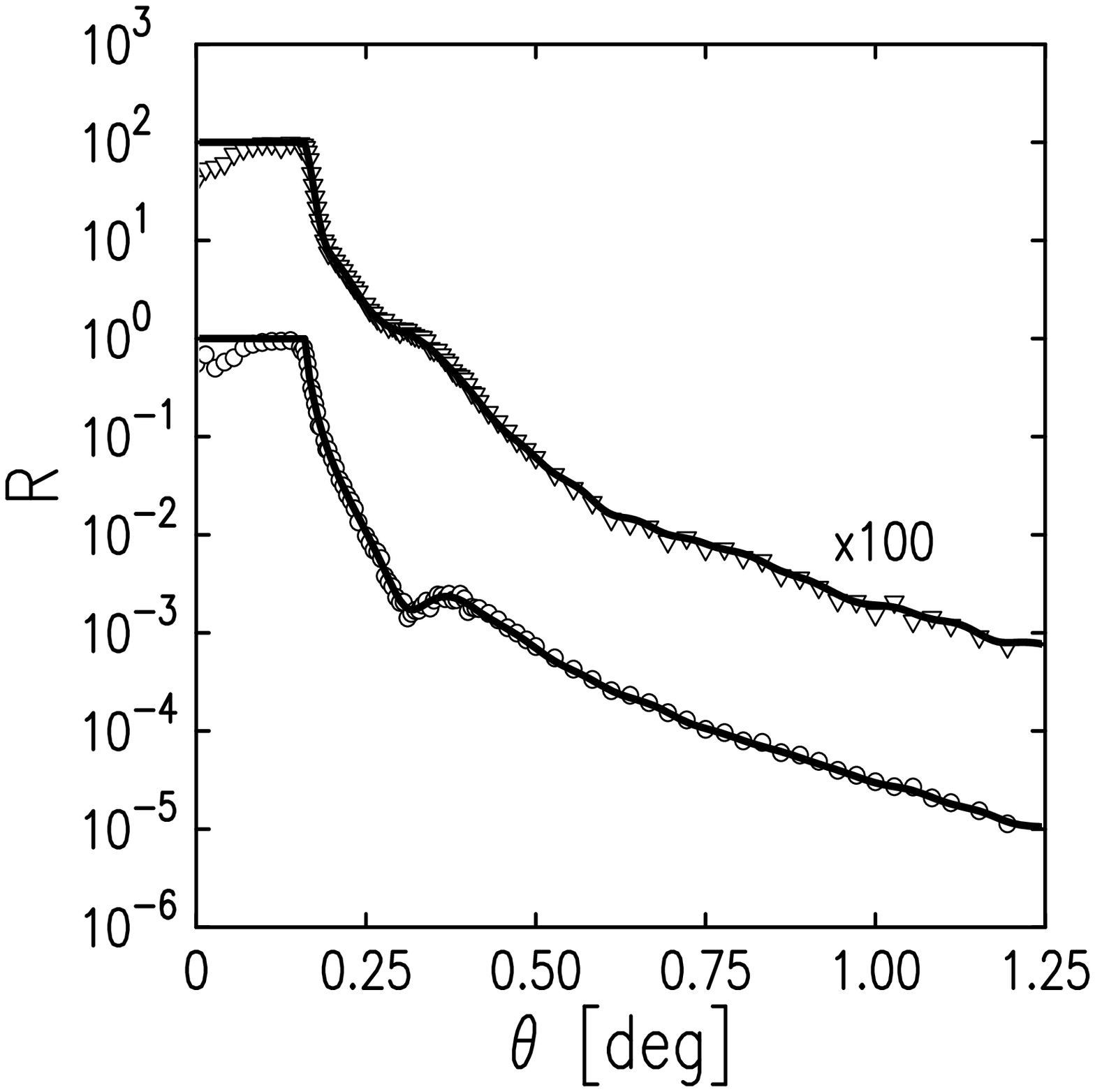, width=0.65\textwidth}

Figure 3. Coefficient of reflection of the air/SM-30 hydrosol
interface for the (circles) pure surface and (triangles) surface with lipid film. The solid lines are simulation by the
method described in [7].
\end{figure}

Figures 3 and 4 show the coefficient of reflection $R$ from the air/hydrosol interface as a function of the
glancing angle $\theta$. Oscillations of the coefficient of reflection at $\theta>0.18\,^{\circ}$ are due to the inhomogeneous depth
distribution of particles in the surface transient layer.

Using the experimental angular dependencies of the coefficients of reflection $R$ (Figs. 3 and 4), we reconstructed the profiles of the
polarizability of the substance $\delta =  {\rm Re}(1-\epsilon)$ (where $\epsilon$ is the complex dielectric constant) with the method
proposed in [7]. These dependencies are rather complicated, but similar for all of the systems (Fig. 5). Variations in the profile of the
polarizability are attributed to spatial inhomogeneities in the distribution of SiO$_2$ and their length is naturally
related to the size of silica particles.

\begin{figure}
\hspace{0.5in}
\epsfig{file=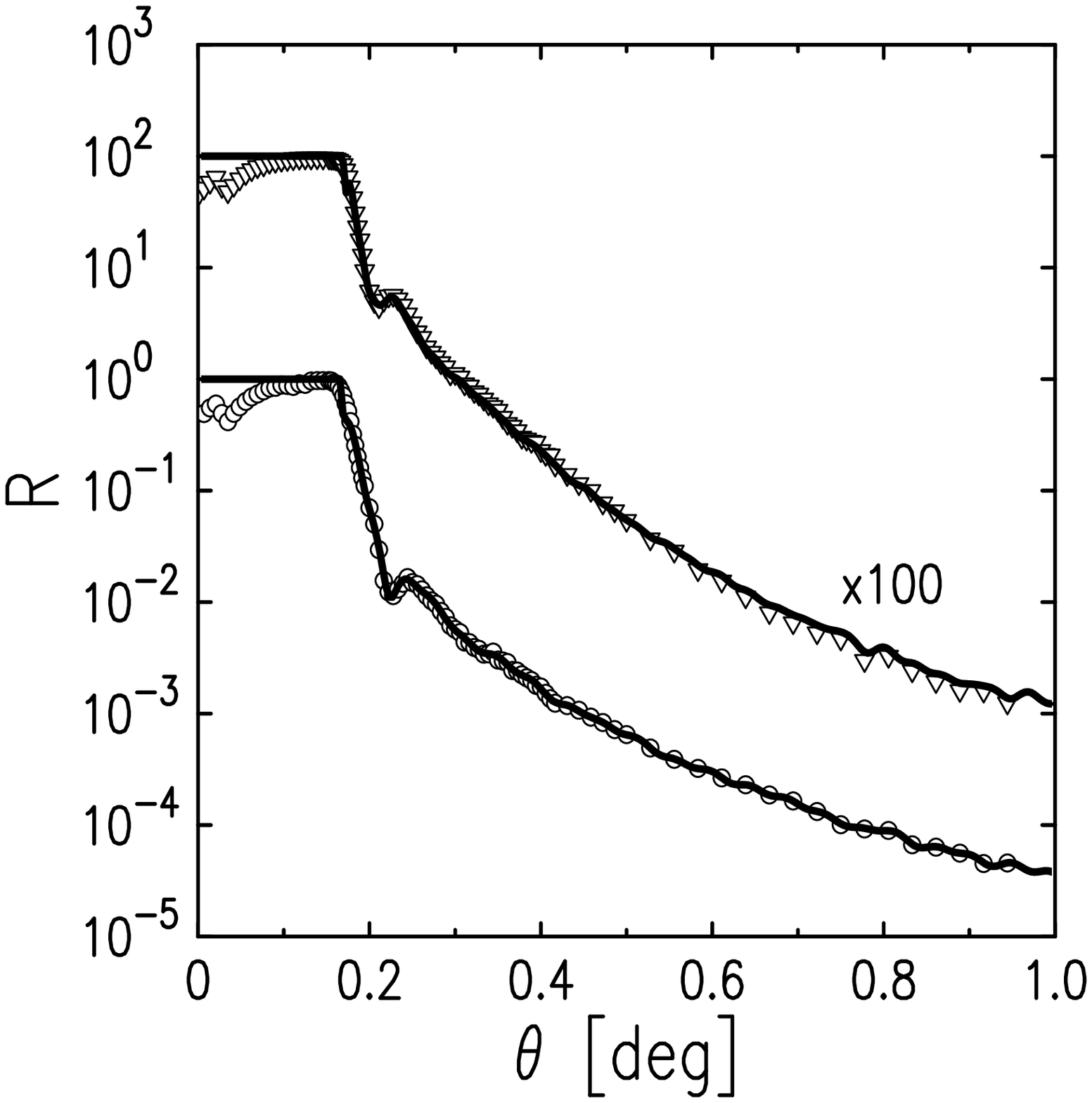, width=0.65\textwidth}

Figure 4. Coefficient of reflection of the air/TM-50 hydrosol interface for the (circles) pure surface and (triangles) 
surface with lipid film. The solid lines are simulation by the method described in [7].
\end{figure}

\begin{figure}
\hspace{0.5in}
\epsfig{file=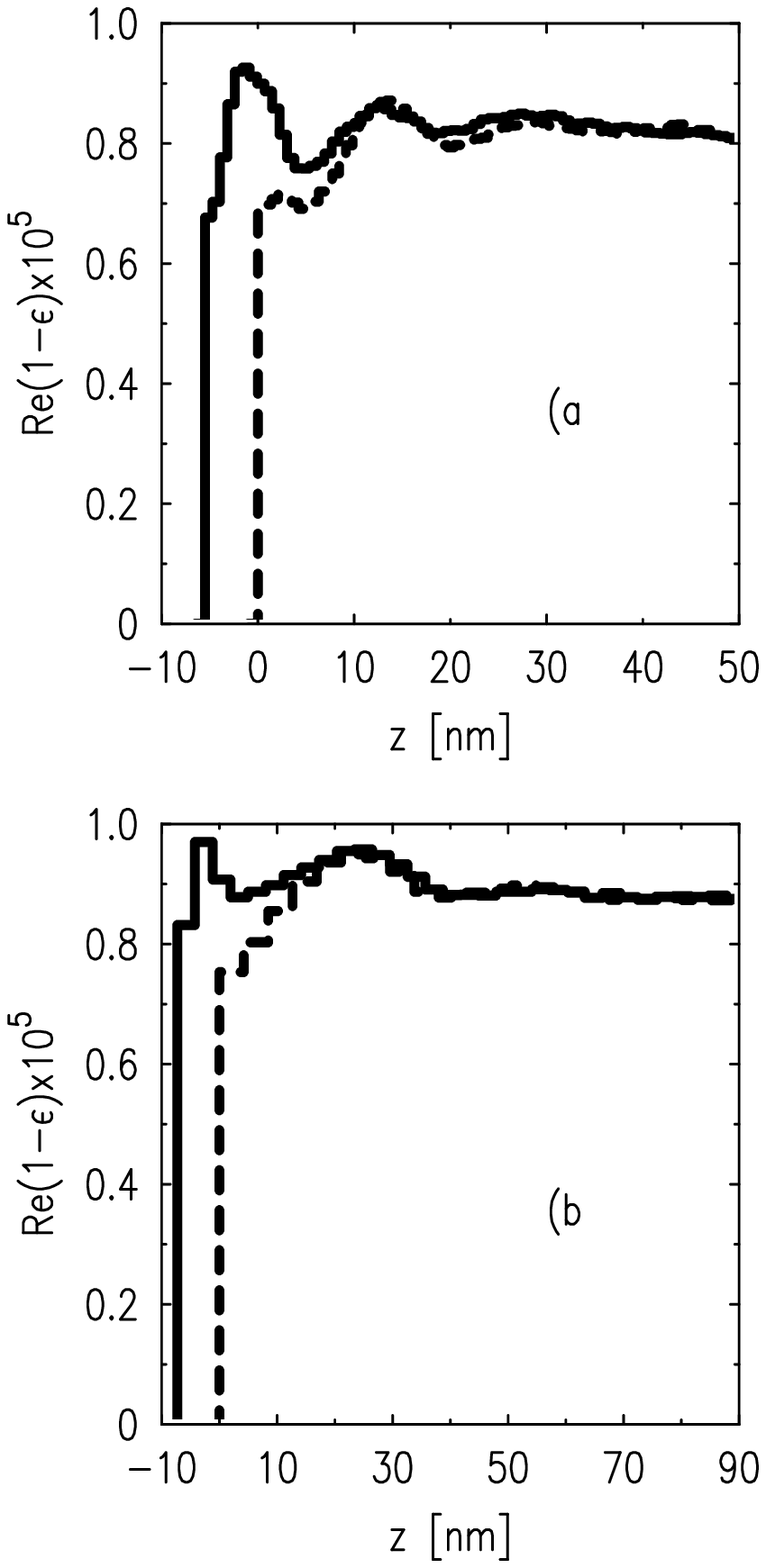, width=0.65\textwidth}

Figure 5. Reconstructed profiles of the real part of the complex dielectric constant across the hydrosol/air interface
(along the z axis) for (a) SM-30 and (b) TM-50 hydrosol. The dashed lines are for the pure hydrosol/air interface.
The solid lines are for the interface with the adsorbed DSPC layer. For pure water, $\delta_w \approx 0.75 \times 10^{-5}$.
\end{figure}

In the case of a pure surface of hydrosol, the most pronounced feature is a dense layer of SiO$_2$ nanoparticles spaced at (SM-30) 15
and (TM-50) 25 nm from the interface with air (Fig. 5), in agreement with [1].
The second, weakly pronounced layer of nanoparticles at (SM-30) 30 and (TM-50) 55 nm from the surface is
also observed. The reconstructed profile indicates that the separation region of silica sol is more extended (by
more than a factor of 2) than that estimated in [1]. The determination of the structure of the transient layer is
more accurate, probably because the monochromaticity of radiation in our experiments is one or two orders
of magnitude higher than that in the experiment reported in [1].

The formation of the first dense layer of nanoparticles at the air/hydrosol interface is equivalent to the
appearance of an additional interface at which the difference between the potentials of electric image forces
for cations and anion particles in the volume of the sol. This leads to the formation of a second layer of
nanoparticles deep inside the sol.

DSPC molecules are hydrated in the aqueous medium and form a strongly polarized hydrophilic
surface [8], which leads to a sharp change in the distribution of SiO$_2$ particles in the transient layer. The
maximum concentration of SiO$_2$ particles for SM-30 hydrosol is reached in the layer immediately adjacent
to the lipid membrane (the polarizability of the membrane is $\sim \delta_w$ [2]) and is higher than the bulk value by a
factor of $(\delta_{max}-\delta_w)/(\delta_b-\delta_w)\approx 2$, where $\delta_w\approx 0.75\cdot10^{-5}$ and $\delta_b \approx 1.15\delta_w$ are the polarizabilities of water and hydrosol, respectively. The maximum polarizability in the transient layer is $\delta_{max}\approx 0.95\cdot10^{-5}$
 (see Fig. 5). In the case of TM-50 hydrosol, we also observed an increase in the surface density of silica
particles 10 nm in diameter, which are apparently impurity.

Thus, silica particles are concentrated at the interface with the phospholipid wall. In this case, the surface 
concentration of nanoparticles, which is characterized by the ratio $(\delta_{max}-\delta_w)/(\delta_{min}-\delta_w)$,
 where $\delta_{min}$ is the polarizability on the pure surface of hydrosol, increases by
more than an order of magnitude. This process should be accompanied by the redistribution of the positive
charge of Na$^+$ cations in the double electric surface layer. The capability of sodium ions to penetrate into
phospholipid membranes and, thus, to create a positive surface potential, which evidently induces the
condensation of negatively charged nanoparticles (their aggregation with the membrane), was discussed
in [9, 10]. In this process, the volume of the sol is a reservoir for Na$^+$.

To summarize, the structure of the transient layer at the interface of the aqueous solution of silica 
nanoparticles has been studied by the X-ray reflectometry method. The reconstructed depth profile of the real
part of the dielectric permittivity indicates the presence of a complicated structure consisting of several layers
of nanoparticles with the thickness that is at least twice as large as the previously estimated value. According to
our data, the adsorption of 1,2-distearoyl-sn-glycero-3-phosphocholine molecules at the hydrosol/air
interface is accompanied by a significant increase in the surface concentration of silica nanoparticles, i.e.,
by their condensation. This phenomenon can be qualitatively explained by the formation of the positive surface 
potential due to the penetration and accumulation of Na$^+$ cations in the phospholipid membrane.

\end{document}